\documentclass[12pt,preprint]{aastex}
\usepackage{graphicx}

\shorttitle{Hot Plasma Diagnostics from Coronal Imaging}
\shortauthors{Testa \& Reale}
\usepackage{color} 
\usepackage{amsmath}           
\usepackage{amsfonts}

\def \arcsec {\hbox{$^{\prime\prime}$}}

\def \hinode   {{\em Hinode}}

\def \eis  {{\sc EIS}}

\def \ov     {O\,{\sc v}}

\def \feix    {Fe\,{\sc ix}}
\def \fex     {Fe\,{\sc x}}
\def \fexi    {Fe\,{\sc xi}}
\def \fexii   {Fe\,{\sc xii}}

\def \fexviii {Fe\,{\sc xviii}}

\def \s9     {S\,{\sc ix}}

\def \caxvii  {Ca\,{\sc xvii}}

\def\ion[#1 #2]{#1\,{\sc #2}}
\def\densr[#1 #2]{10$^{#1}$\hskip 1pt{--}\hskip .5pt{10$^{#2}$}\hskip 1.5pt{cm$^{-3}$}}
\def\fl[#1 #2]{{#1}$\pm${#2}}
\def\orb[#1 #2]{{$#1^{#2}$}}
\def\ls[#1 #2]{{$^{#1}${#2}}}
\def\tm[#1 #2 #3]{{$^{#1}${#2}$_{#3}$}}

\newcounter{ion}

\begin{document}

\title{Hinode/EIS spectroscopic validation of very hot plasma imaged 
  with Solar Dynamics Observatory in non-flaring active region cores }

\author{Paola Testa$^{1}$}
\email{ptesta@cfa.harvard.edu}
\author{Fabio Reale$^{2,3}$}

\affil{$^1$ Smithsonian Astrophysical Observatory,60 Garden street, MS
  58, Cambridge, MA 02138, USA} 
\affil{$^2$Dipartimento di Fisica, Universit\`a di Palermo, Piazza del Parlamento 1,
90134, Italy}
\affil{$^3$INAF-Osservatorio Astronomico di Palermo, Piazza del Parlamento 1, 
        90134 Palermo, Italy}

\begin{abstract}

We use coronal imaging observations with {\em SDO/AIA},
and {\em Hinode/EIS} spectral data, to explore the potential of 
narrow band EUV imaging data for diagnosing the presence of hot 
($T \gtrsim 5$~MK) coronal plasma in active regions.
We analyze observations of two active regions (AR 11281, AR 11289) 
with simultaneous {\em AIA}  imaging, and {\em EIS} spectral data, 
including the \caxvii\ line (at $192.8$\AA) which is one of the few 
lines in the {\em EIS} spectral bands sensitive to hot coronal 
plasma even outside flares.
After careful coalignment of the imaging and spectral data, we compare
the morphology in a 3 color image combining the 171, 335, and 94\AA\ 
{\em AIA} spectral bands, with the image obtained for \caxvii\ emission
from the analysis of {\em EIS} spectra.
We find that in the selected active regions the \caxvii\ emission is strong 
only in very limited areas, showing striking similarities with the features 
bright in the 94\AA\ (and 335\AA) {\em AIA} channels and weak in the 
171\AA\ band.  
We conclude that {\em AIA} imaging observations of the solar corona
can be used to track hot plasma (6-8~MK), and so to study its 
spatial variability and temporal evolution at high 
spatial and temporal resolution.

\end{abstract}


\section{Introduction}
\label{s:intro}

In spite of remarkable progress, and although it is well established that 
the ultimate energy source is the coronal magnetic field, the question of 
how the magnetic energy is transformed to heat the coronal plasma is still 
to be solved \citep[e.g.,][]{Klimchuk06,Reale10}. 
Among the several aspects debated in the literature one important issue 
is whether the heating is released gradually and continuously or in the 
form of discrete, rapid and intense pulses. Several physical and technical 
reasons inhibit a direct answer, e.g., the efficient thermal 
conduction along the magnetic field lines, the low emission from the 
heating site, and limited spatial and temporal resolution \citep{Klimchuk06}. 
The question is made even more difficult by the increasing evidence for 
an extremely fine-structured confined corona: the magnetic flux tubes 
that confine the bright coronal plasma are collections of a multitude of 
thin strands, of expected cross-section below the resolution of past and current 
coronal telescopes \citep{Gomez93,Vekstein09,Guarrasi10}. 
Finding direct signatures with current instruments is therefore not feasible 
for these inherently elusive mechanisms that occur on very small spatial scales.
Indirect diagnostics are thus the only viable way at present. A feature that may 
strongly discriminate between gradual and impulsive heating release is 
the presence or absence of plasma at very high temperature. If we consider 
an active region, in order to keep the coronal loops at the typical temperature 
of 2-3 MK, we expect that occasional heat pulses should be so strong as 
to temporarily heat the plasma temperature to several MK, even above 10. 
Since we expect these heating episodes to be spread over all coronal 
structures, we might also search for a consequent minor but steady very 
hot plasma ($\sim 6 - 10$~MK) component. This might appear either as a 
hot tail or as a secondary peak to the hot side of the plasma emission 
measure distribution. Increasing evidence for this minor hot component 
has been collected in recent times, both with filter ratios or differential 
emission measure (DEM) reconstruction from imaging broad-band instruments 
(XRT, e.g., \citealt{Reale09,Reale09b,McTiernan09,Schmelz09a}) and with 
spectroscopic line analysis (single line or DEM reconstruction, e.g.,
\citealt{Ko09,Patsourakos09,Shestov10,Sylwester10}), 
but it is still not conclusive for various reasons.

Spectroscopy would be ideal to reveal hot plasma but the moderate spatial 
and temporal coverage inhibits systematic campaigns and makes any 
detection or non-detection partial and limited \citep[e.g.,][]{Warren11}. 
Also, DEM reconstructions typically suffer from severe limitations due to
several factors such as the small number of line fluxes available in a 
narrow range of temperature, uncertainties in the element abundances,
and, also in atomic data \citep[e.g.,][]{Testa11}. This makes DEM reconstruction 
reliable on a large temperature scale at most, but much less on the 
details (e.g., \citealt{Landi12}; Testa et al.\ 2012, in preparation).
It becomes then appealing and reasonable to search for methods to 
discover hot plasma components with simple diagnostics and without 
addressing DEM reconstruction methods.

Recently \cite{Reale11} have used a different approach to infer
the widespread presence of 6-8~MK plasma in the core of an active 
region. A model of multi-stranded pulse-heated coronal loop 
had predicted that if any very hot component were present, it would 
be characterized by spatial distribution of its emission significantly 
more sparse compared to the 2-3~MK emission \citep{Guarrasi10}. 
The Atmospheric Imaging Assembly (AIA; \citealt{Lemen12}) on board 
the {\em Solar  Dynamics Observatory} (SDO) is based on 
normal-incidence optics and is equipped with 6 narrow-band filters 
in the EUV band (94-335\AA) that contain a few strong spectral lines 
that sample the solar corona in a wide temperature range, approximately 
between 0.5 and 10~MK. 
The instrument continuously monitors the full-disk corona with high 
cadence ($\lesssim 12$s) and a pixel size of $\sim 0.6$~arcsec. 
One of the filter, the 94\AA\ channel, includes a highly ionized Fe line, \fexviii, 
that is sensitive to plasma at temperature in the range 6-8~MK. 
\cite{Reale11} show that the bright 94\AA\ emission in an active 
region core matches very well the filamented aspect predicted in the 
modeling paper. However, the 94\AA\ passband nominally includes other 
spectral lines that make it sensitive also to plasma emitting in a 
range around 1~MK \citep{Boerner12}. The radically different morphology of the image
obtained in the 171\AA\ channel, containing a very strong \feix\ 
line sensitive to plasma at $\sim 1$~MK, made the authors conclude 
that most of the plasma detected in the 94\AA\ channel was in the 
hot branch. However, the lack of spectral resolution in the {\em AIA} 
observations leaves room to unknown line contributions that make 
the result not unequivocal.

Therefore, conclusive evidence whether the 94\AA\ bright emission in 
the core of active regions can be ascribed to hot plasma is expected to 
come from comparison with spectroscopic data. The Extreme-ultraviolet 
Imaging Spectrometer  ({\em EIS}) on board {\em Hinode} \citep{Culhane07} 
is sensitive in the EUV band (in the 171-212\AA\ and 245-291\AA\ 
spectral ranges) and is able to provide images (spectroheliograms) in 
spectral lines thanks to rastering programs that take few minutes to 
hours, depending on the target and field of view (FOV). 
In spite of the moderate spatial and temporal resolution, this 
instrument is appropriate for our task, since it has imaging capabilities,
and its spectral range includes a \caxvii\ line that is emitted 
by plasma at 6-8~MK \citep{Patsourakos09,Ko09,Warren11}. 
The scope of this work is to validate the detection 
of very hot plasma in active regions through the direct comparison 
of the morphology simultaneously observed with {\em Hinode/EIS }in 
the \caxvii\ line and with {\em SDO/AIA} in the 94\AA\ channel. 
Finding that well-defined bright structures have the same appearance 
in both observations would be a very strong evidence that the bright 
94\AA\ emission in the AR cores really is due to plasma at 6-8~MK.

The task that we have tackled in this work is not so easy for various 
reasons: among the active regions that in the 94\AA\ {\em AIA} channel 
show the presence of hot plasma, we have to find examples observed 
at the same time with {\em EIS} in the band including the \caxvii\ line 
and including the hot plasma region in its FOV.  
Furthermore, the emission in the \caxvii\ line is not trivial to extract, 
because of blending with other lines, and consequently the deblended
emission can be trusted only in regions where it is strong \citep{Ko09}. 
Finally, also the coalignment of the obtained {\em EIS} \caxvii\ image
with the {\em AIA} images is non-trivial. 

In Section~\ref{s:method} we describe the data selection, co-alignment
procedure, analysis methods, and results. 
We discuss our findings and draw our conclusions in Section~\ref{s:conclusions}.

\section{Data Selection, Analysis Methods and Results}
\label{s:method}

We searched for active region observations with simultaneous {\em SDO/AIA} 
imaging, and {\em Hinode/EIS} spectral data, including the {\em EIS} short
wavelength range (171-212\AA), where the hot \caxvii\ emission
line lies ($\approx 192.85$\AA).
We selected two active regions, AR 11281 and AR 11289, observed close to 
disk center in September 2011.

The selected {\em EIS} rasters are characterized by field of view of
120\arcsec$\times$160\arcsec, slit width of 2\arcsec, and exposure time 
of 60~s at each step (study acronym Atlas\_60).  The images are built up 
by stepping the 2\arcsec\ slit from solar west to east over a $\sim 1$~hr 
period. The study takes full spectra on both the \eis\ detectors from 
171-212\AA\ and 245-291\AA. 
The {\em EIS} observations of AR 11281 have a start time of 2011 September
02 at 23:32UT, whereas the observations of AR 11289 started on 2011
September 13 at 10:37UT.
The \eis\ data are processed with the eis\_prep routine available in 
SolarSoft to remove the CCD dark current, cosmic-ray strikes on the 
CCD, and take into account hot, warm, and dusty pixels. In addition, 
the radiometric calibration is applied to convert the data from 
photon events to physical units.  The \eis\ routine eis\_ccd\_offset is 
also used to correct for the CCD offset. 
We selected simultaneous {\em AIA} data in the 171\AA, 335\AA, 94\AA, and
193\AA\ passbands, and processed the level 1 data with the aia\_prep routine,
which performs image registration (co-alignment, and adjustments for the 
different plate scales and roll angle) and is also available as a part of SolarSoft.

\begin{figure}[!ht]
\centerline{\includegraphics[scale=0.3]{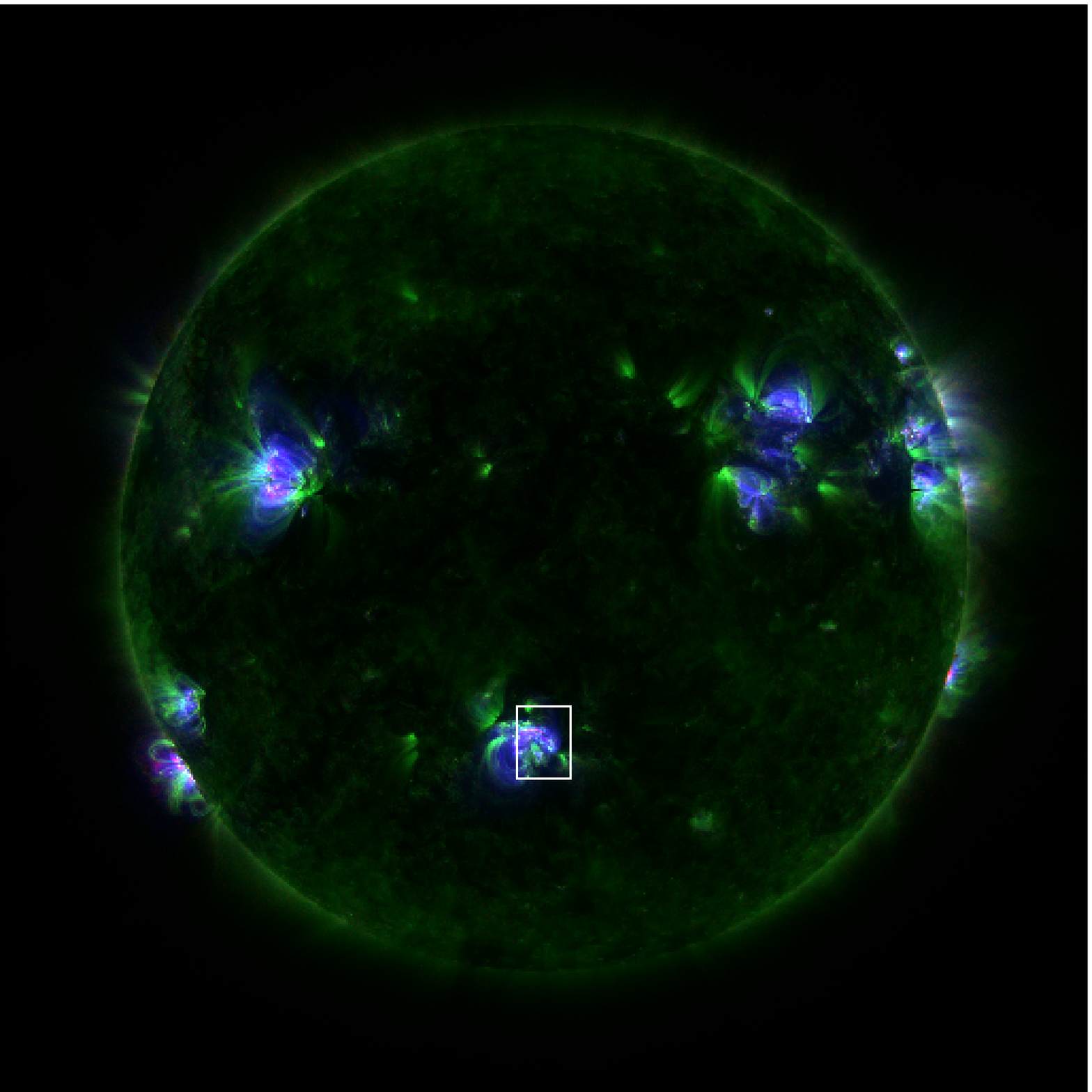}\hspace{0.2cm}
  \includegraphics[scale=0.3]{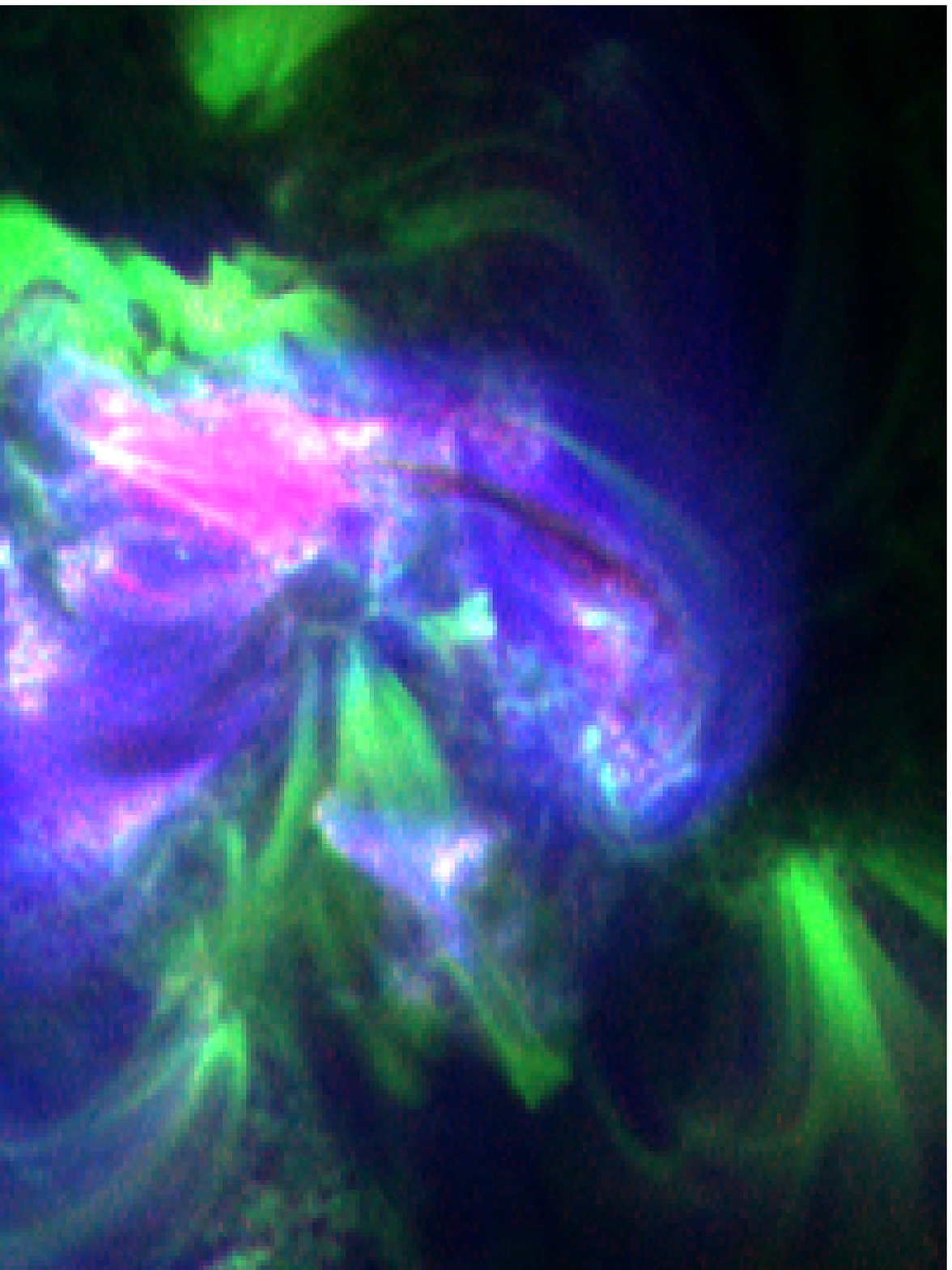}\hspace{0.2cm}
  \includegraphics[scale=0.3]{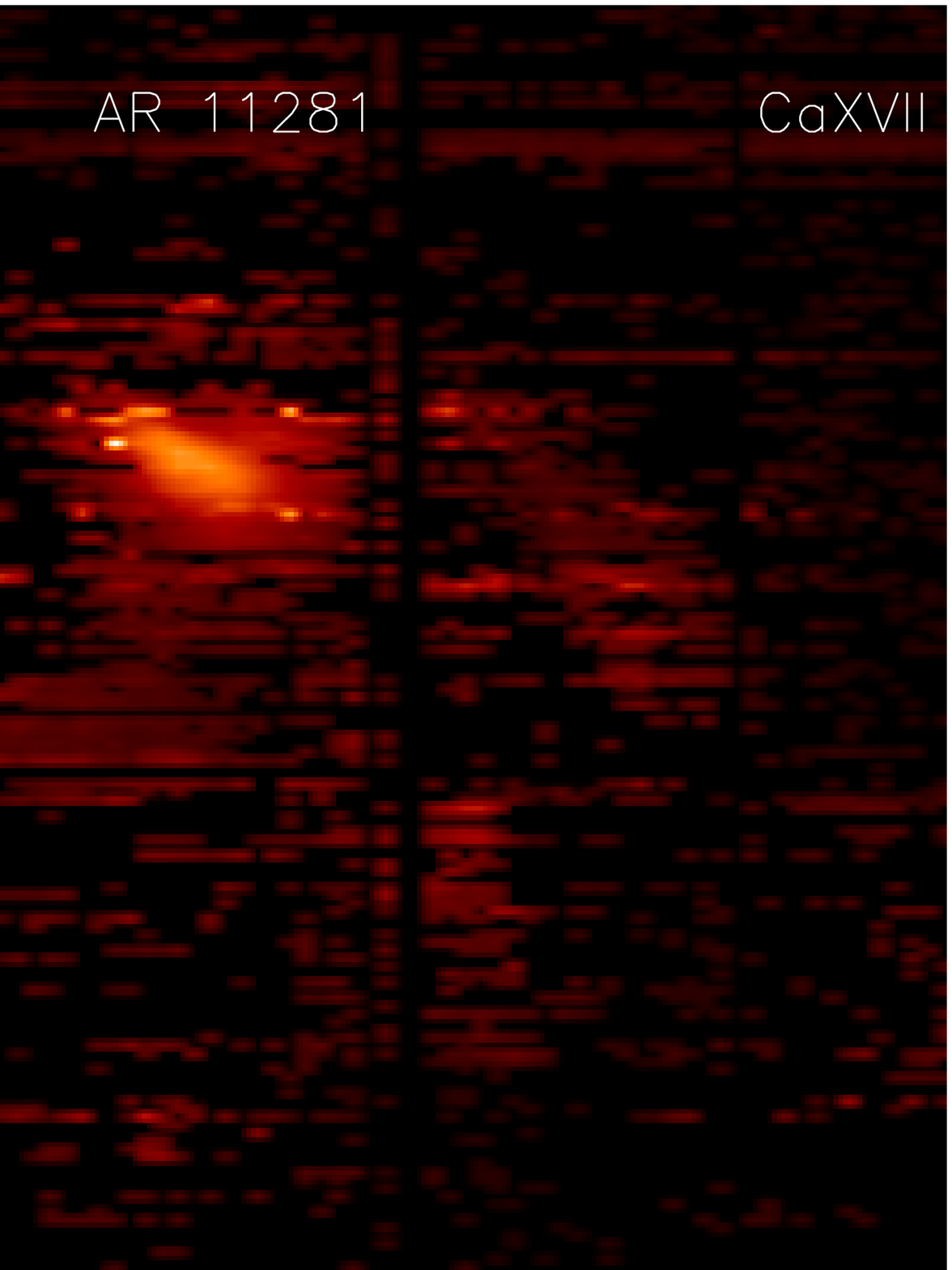}}\vspace{0.2cm}
\centerline{\includegraphics[scale=0.3]{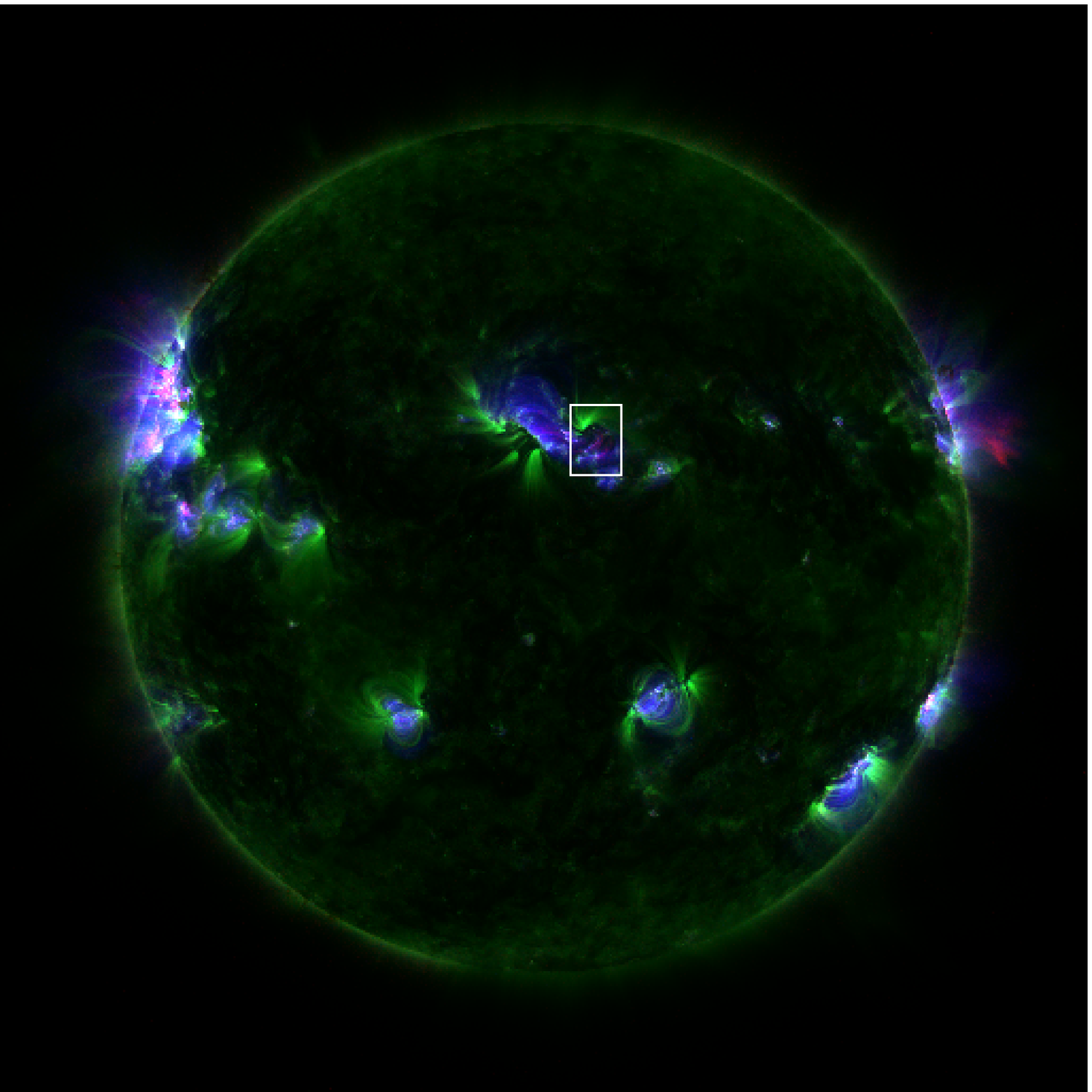}\hspace{0.2cm}
  \includegraphics[scale=0.3]{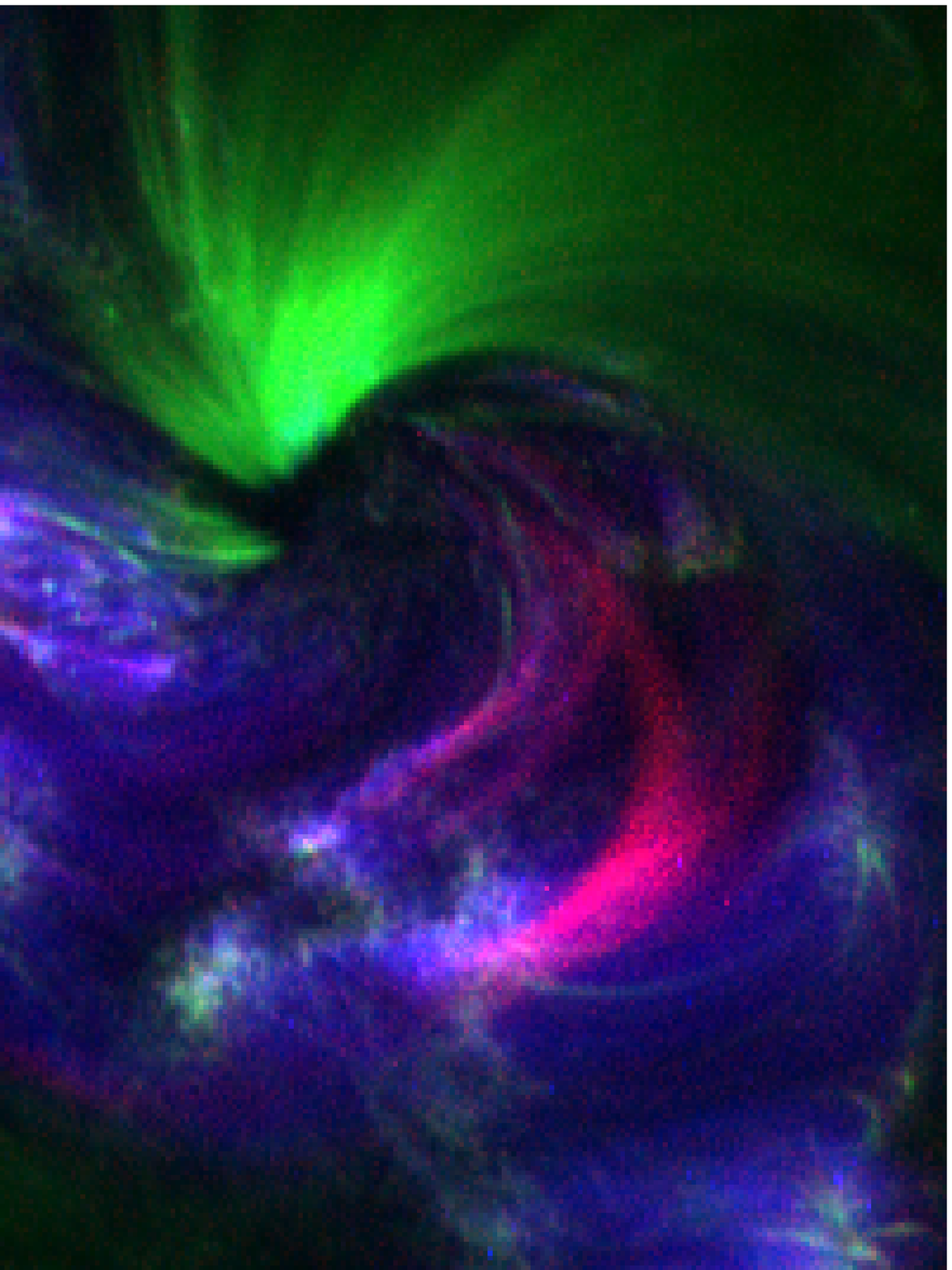}\hspace{0.2cm}
  \includegraphics[scale=0.3]{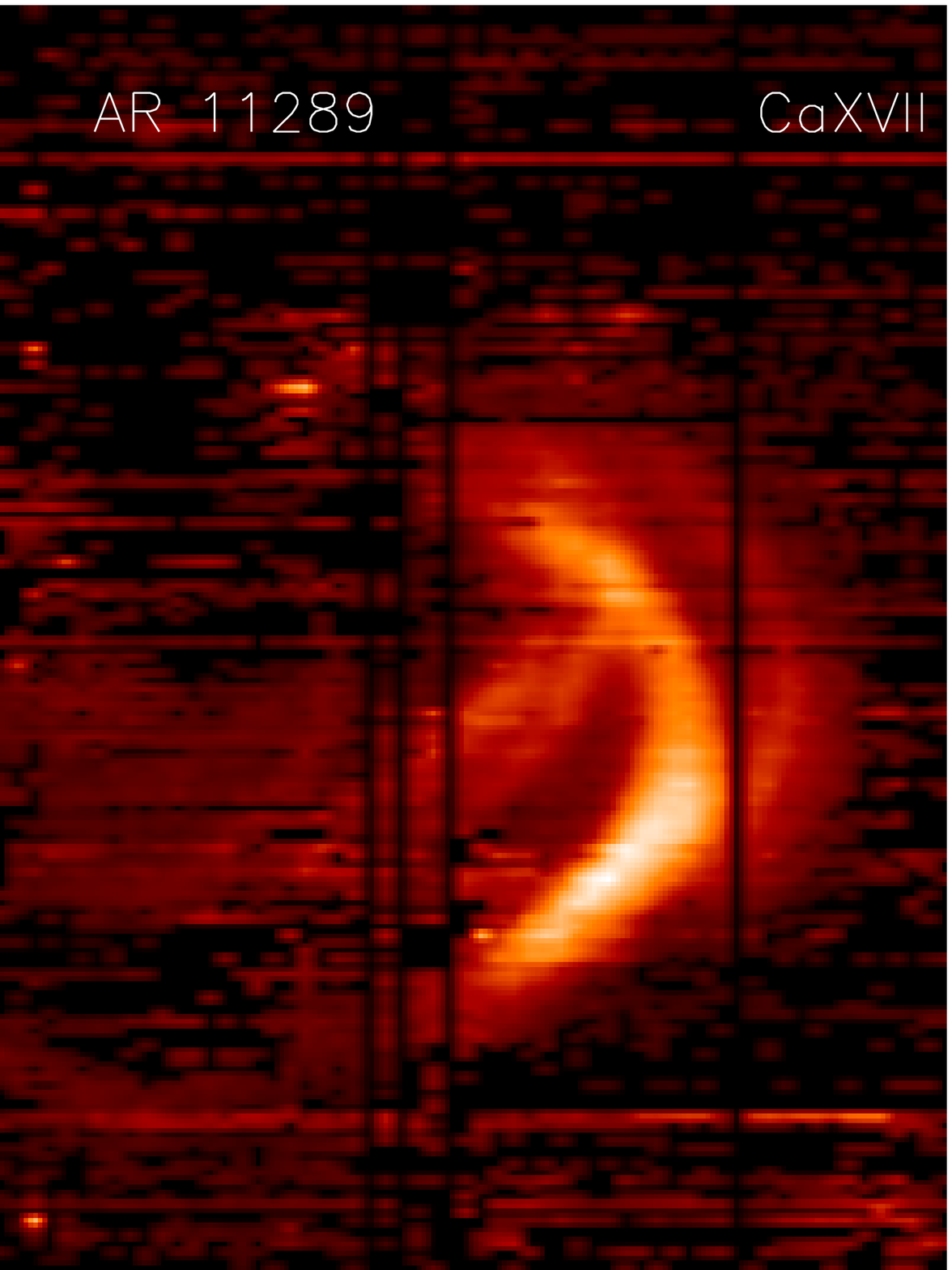}}\vspace{0.2cm}
\caption{{\em SDO/AIA} and {\em Hinode/EIS} observations of AR 11281 
  ({\em top row}), and AR 11289 ({\em bottom row}). We show the AIA full 
  disk data at the time of the beginning of the {\em EIS} observations, 
  i.e., 2011 September 2 around 23:32UT and 2011 September 13 around 
  10:37 for AR 11281 and AR 11289 respectively. 
  {\em Left}: {\em SDO/AIA} 3 color full disk image combining {\em AIA} 
  observations in the 171\AA\ (green), 335\AA\ (blue), and 94\AA\ (red) 
  channels. The field of view of the EIS observations (120~arcsec $\times$ 
  160~arcsec) is shown. {\em Center:} AIA 3 color image for the {\em EIS} 
  FOV (94\AA, 171\AA, and 335\AA\ correspond to red, green, and blue 
  respectively, as in the left panels) created by building for each channel a 
  composite image, where for each vertical slice in the {\em EIS} FOV we 
  select the {\em AIA} data closest in time. 
  {\em Right:} Image of \caxvii\ 192.85\AA\ emission from
  the {\em EIS} raster spectral data. 
  \label{fig:AIA_EIS}} 
\end{figure}

In Figure~\ref{fig:AIA_EIS} (left panels) we show the 3 color full disk {\em AIA} 
images combining the 171\AA, 335\AA, and 94\AA\ intensities (green, blue, 
and red respectively), at the start time of the {\em EIS} observations. The field of 
view of the EIS observations (120~arcsec $\times$ 160~arcsec) is marked on
the full disk {\em AIA} images.

We coaligned the datasets from the two instruments by applying a standard 
cross-correlation routine (tr\_get\_disp.pro which is part of the IDL 
SolarSoftware package), to the {\em EIS} images of the \fexii\ 193\AA\ 
emission obtained from fitting the {\em EIS} spectra, and the 193\AA\ 
{\em AIA} synthetic rasters (composite image, hereafter) built as follows:
each vertical stripe is extracted from the AIA image closest in time to 
the time when the {\em EIS} slit was at that location.   
After co-aligning the time series {\em AIA} images in each channel (for the 
1 hour period corresponding to the {\em EIS} observations), and between 
channels, and to the {\em EIS} data, we created {\em AIA} composite 
images in each channel. In Figure~\ref{fig:AIA_EIS} (middle panels) we 
show the 3 color images, obtained by combining the 171\AA, 335\AA, 
and 94\AA, composite images in the {\em EIS} FOV. For each channel the 
intensity range chosen for the color scale covers a factor 40, where the 
maximum is roughly half of the intensity maximum in the channel; the 
color scale is the same for the two datasets.
Most features in the 3 color images appear dominated by a single color, the
moss being a clear exception, as expected. For instance, the fan loops at
the boundary of the AR are predominantly green (i.e., with dominant
emission in the 171\AA\ band), whereas blue (i.e., 335\AA) largely characterizes 
the emission of active region plasma. A few AR features are 
however dominantly pink-red, indicating significant emission in the 94\AA\ 
and 335\AA\ bands, with negligible contribution from the 171\AA\ emission. 
The contribution of different spectral features to the {\em AIA} 94\AA\ channel,
as predicted using the CHIANTI database (\citealt{chianti,chianti6}; the {\em AIA} 
responses as a function of temperature can be obtained using the SolarSoft 
routine aia\_get\_response), are dominated by \fexviii\ emission
at the high temperature end ($\log T[K] \gtrsim 6.5$), and by \fex\ for 
warm plasma ($\log T[K] \sim 6$). However, recent studies support 
the presence of significant additional contributions to the 94\AA\ {\em AIA}
passband missing from current atomic databases (e.g., \citealt{Testa12}; 
also, H.\ Warren, priv.\ comm., has derived an empirical model of the warm
contribution to the 94\AA\ emission by scaling a combination of the 171\AA,
and 193\AA\ emission).

In order to validate the hypothesis that the 94\AA\ emission for the 'pink' 
features in the 3 color image is dominated by \fexviii, i.e., emitted by hot 
plasma, we analyze the {\em EIS} spectra to measure the \caxvii\ emission.
We fit the \caxvii\ lines using the procedure developed by \cite{Ko09}: 
we first fit the \fexi\ lines at $\sim 188$\AA, and then use their intensity to
the model the \fexi\ contribution to the 192\AA\ feature which is a blend of 
\fexi, \ov, and \caxvii.  As discussed by \cite{Ko09} the measurement of the 
\caxvii\ emission from the multi-component fit to the 192\AA\ blend, is 
deemed unreliable when its intensity is lower than 10\% of the total intensity 
of the blend. Therefore we assign zero value to the \caxvii\ intensity of 
those pixels meeting this condition.
In the right panels of Figure~\ref{fig:AIA_EIS} the maps of the \caxvii\ intensity
are shown, and, though somewhat noisy, they present clear evidence of confined
bright features with morphology analogous to the pink features of the {\em AIA}
color image. 

\begin{figure}[!ht]
\centerline{\includegraphics[scale=0.35]{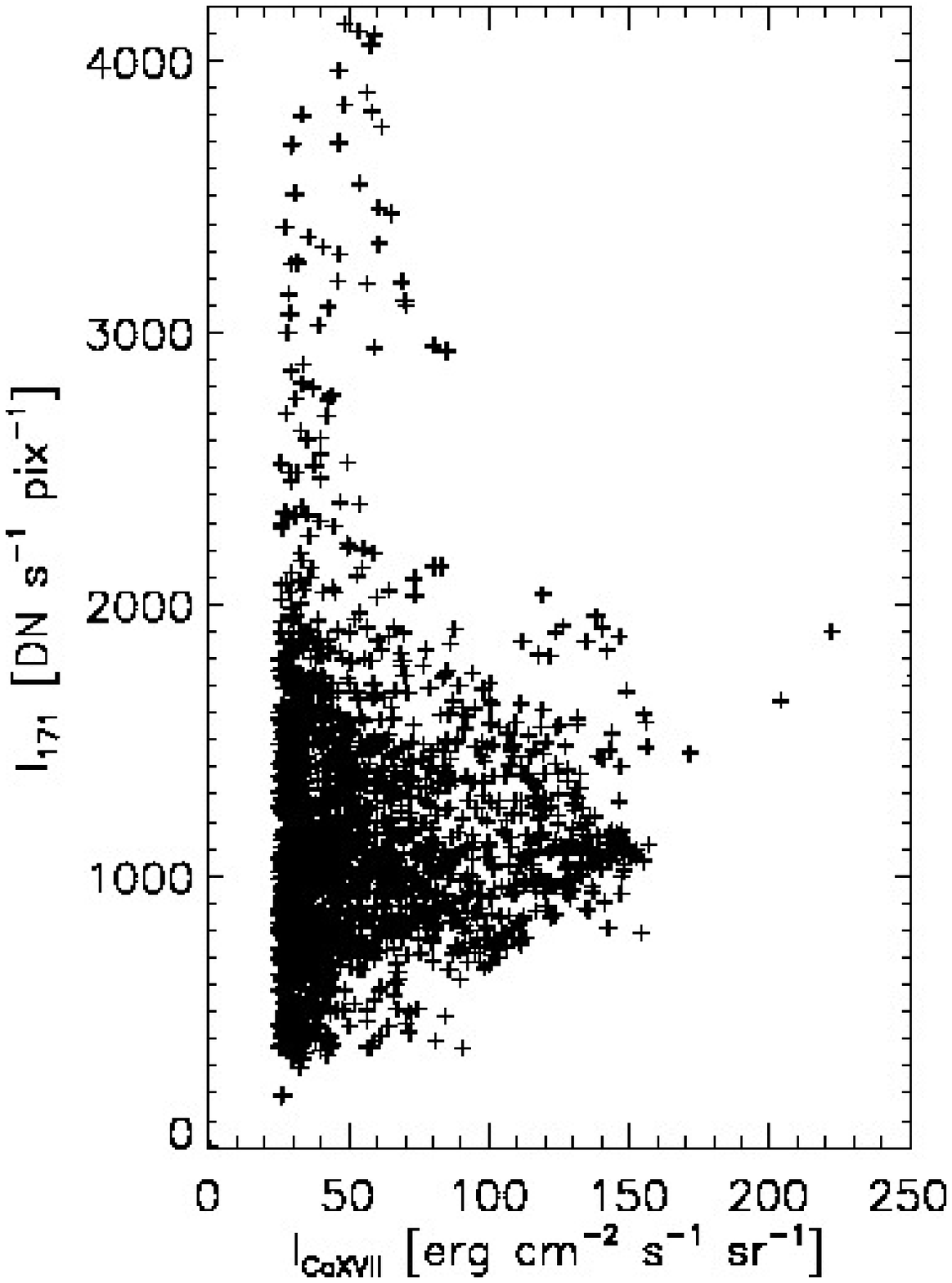}\hspace{-0.2cm}
  \includegraphics[scale=0.35]{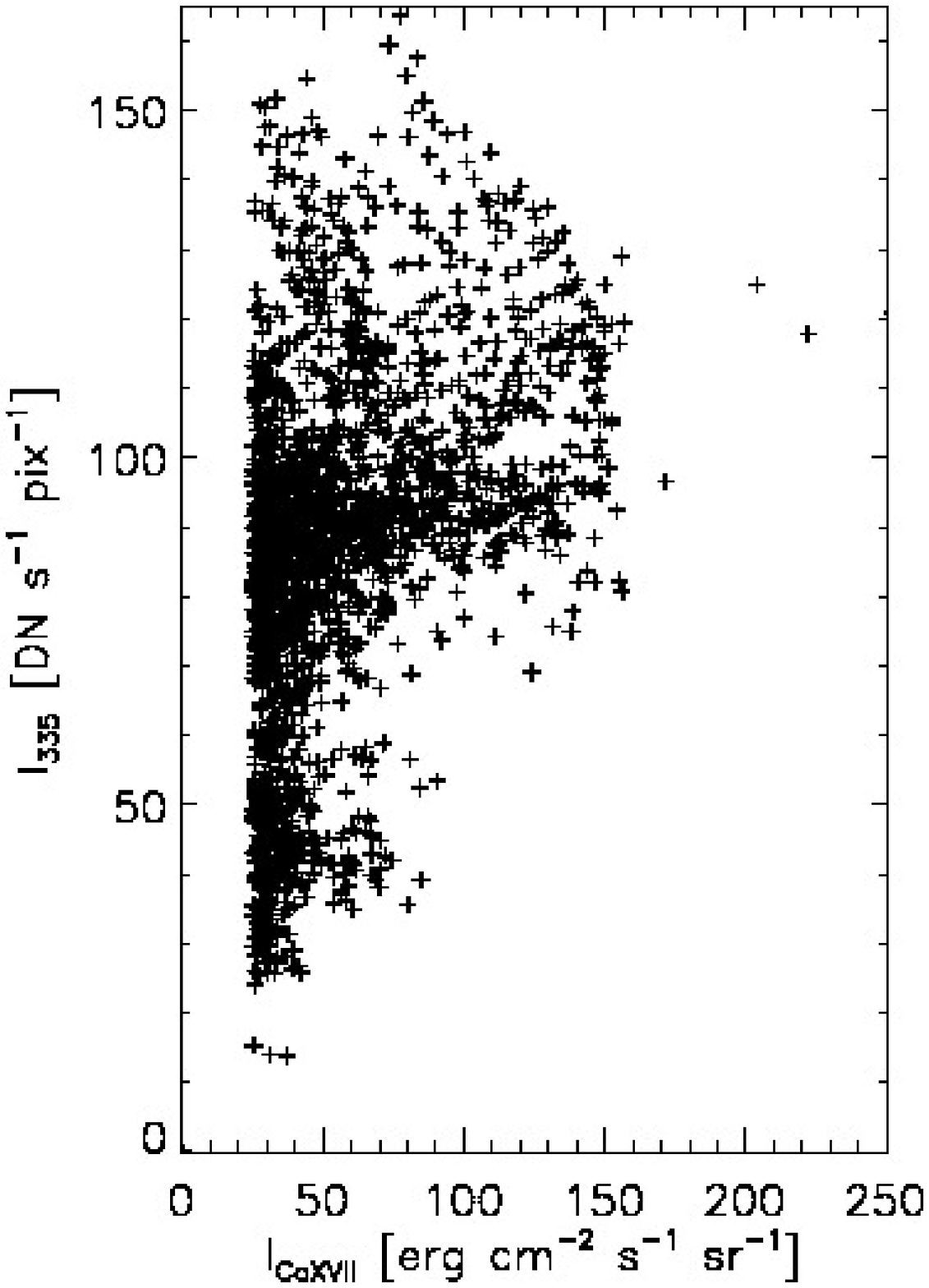}\hspace{-0.2cm}
  \includegraphics[scale=0.35]{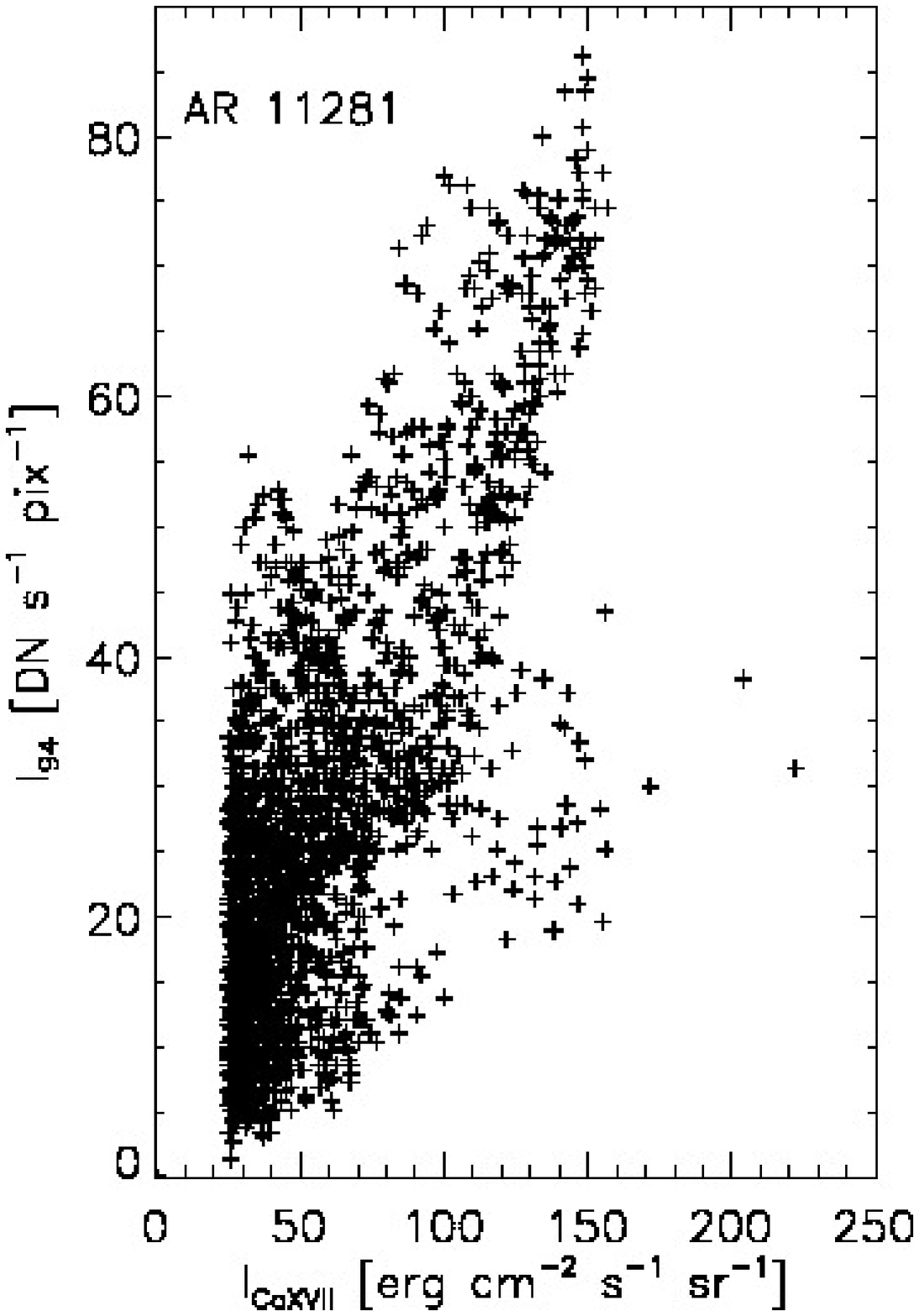}}%\vspace{0.2cm}
\centerline{\includegraphics[scale=0.35]{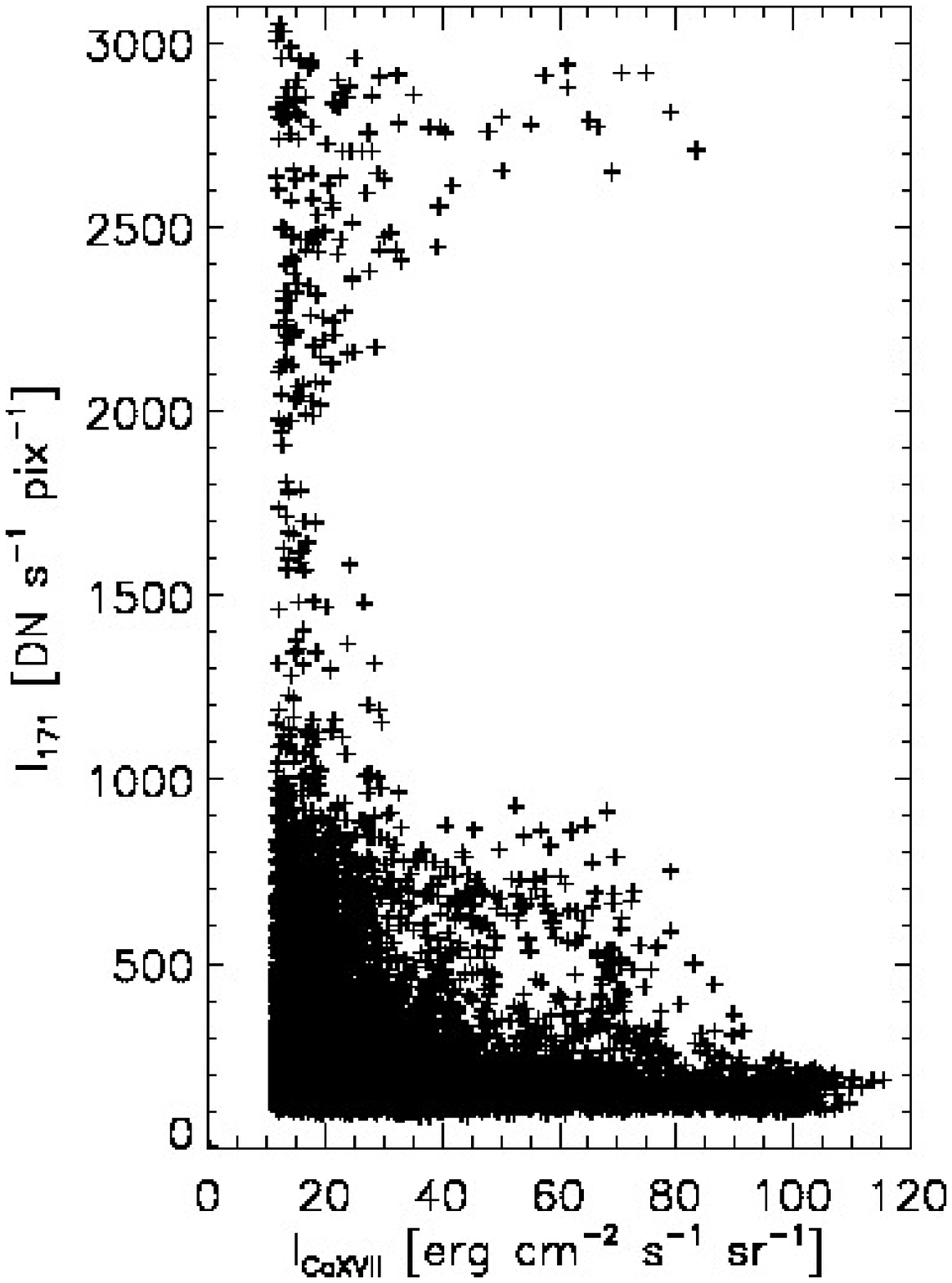}\hspace{-0.2cm}
  \includegraphics[scale=0.35]{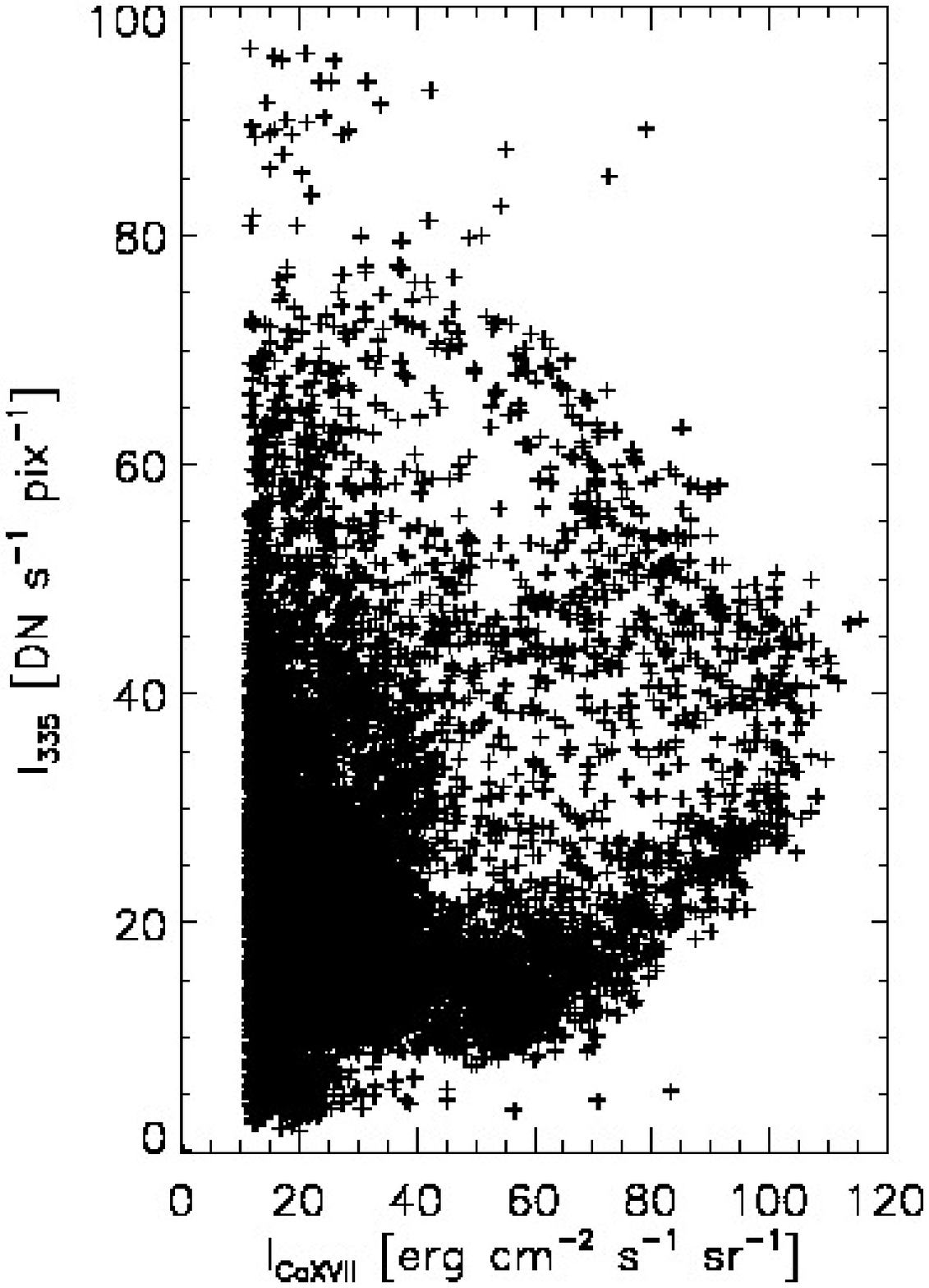}\hspace{-0.2cm}
  \includegraphics[scale=0.35]{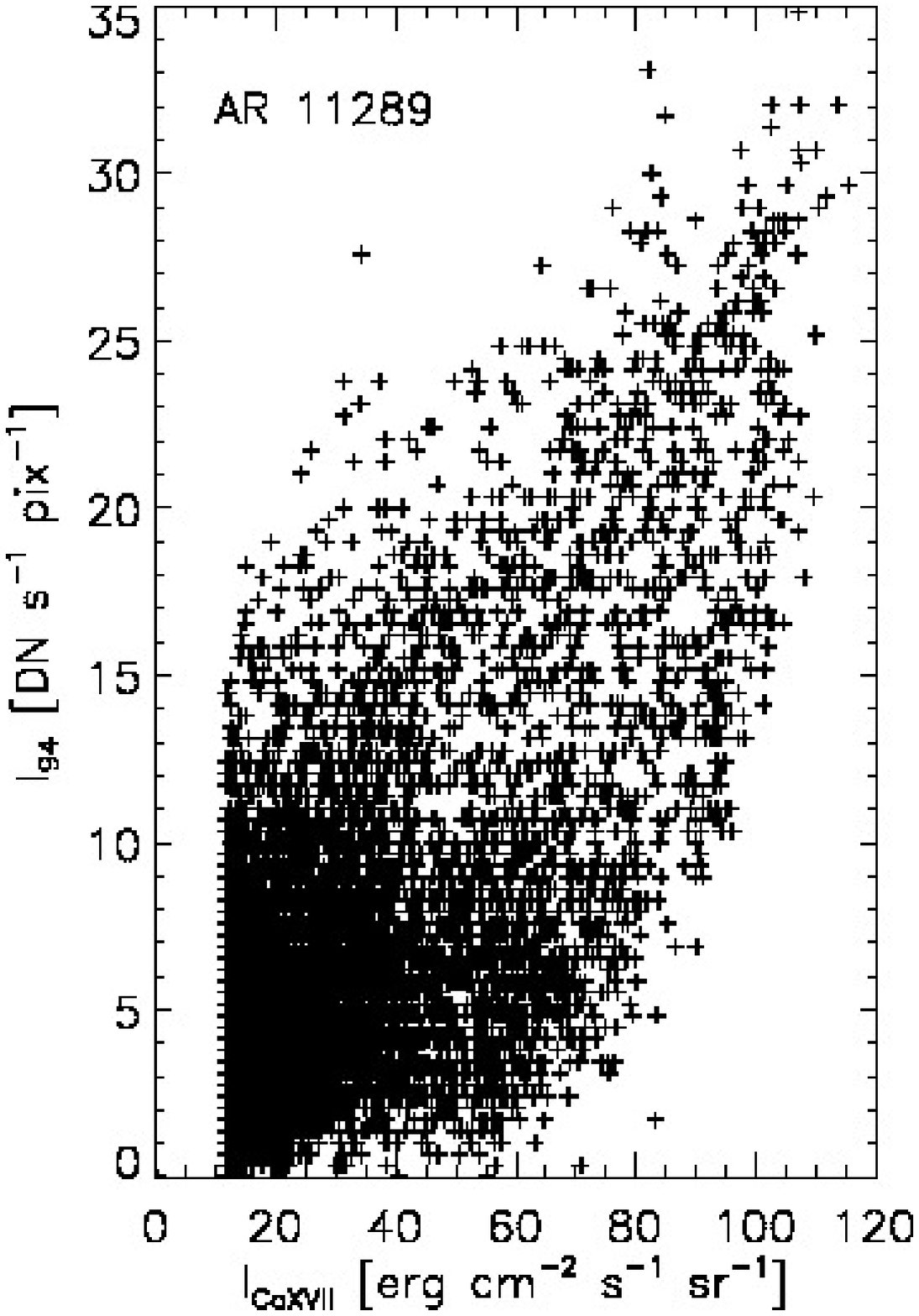}}\vspace{0.2cm}
\caption{Intensity in the AIA 171\AA\ ({\em left}), 335\AA\ ({\em middle}), and 
  94\AA\ ({\em right}) channels (in units of DN~s$^{-1}$~pix$^{-1}$), plotted as a 
  function of the intensity of the \caxvii\ emission line (in units of 
  erg~cm$^{-2}$~s$^{-1}$~sr$^{-1}$). We selected only the pixels where the \caxvii\ 
  intensity was larger than 10\% of the maximum value of \caxvii\ emission
  in the entire {\em EIS} FOV.
  \label{fig:correl}} 
\end{figure}

Further more quantitative support is provided by scatter plots of the 
intensity in each {\em AIA} passband vs.\ the \caxvii\ intensity 
(Figure~\ref{fig:correl}), indicating a clear correlation of the 94\AA\ 
intensity with the \caxvii\ emission, which has no 
clear correlation instead with the two other {\em AIA} bands.

\section{Discussion and conclusions}
\label{s:conclusions}

The determination of the high temperature end ($\log T[K] \gtrsim 6.7$) 
of the plasma thermal distribution is crucial to constraining the coronal
heating \citep[e.g.,][]{Klimchuk06,Reale10}.  In the literature, typical
studies address this issue aiming at a full reconstruction of the emission
measure distribution, based on either spectral or multi-filter 
(broadband/narrowband) imaging data \citep[e.g.,][]{DZM03,Landi09,
Patsourakos09,Brooks09,Warren09,Reale09,Sylwester10,Testa11}. 
However, the determination of the DEM presents significant challenges
because of difficulties inherent to the method (see e.g., 
\citealt{Craig76,McIntosh00,Judge97,Judge10,LandiKlimchuk10,Landi12}, Testa 
et al.\ 2012 in prep.), and limitations of available data, such as their
spatial and temporal coverage (especially for spectral observations),
and temperature diagnostics (especially for imaging data). 

In this Letter we adopted a different approach and investigated the
diagnostic potential of {\em SDO/AIA} imaging observations to diagnose
the hot component ($\log T[K] \gtrsim 6.7$) of the coronal plasma 
distribution.
The comparison of {\em AIA} 3 color images and {\em EIS} \caxvii\ images
shows a highly consistent scenario that confirms the presence of hot 
plasma in many AR cores, as pointed out in \cite{Reale11}.
Figure~\ref{fig:AIA_EIS} shows that the combination of the emission 
in the 171\AA, 335\AA, 
and 94\AA\ {\em AIA} spectral bands is extremely effective in highlighting 
the hot plasma. We validate our findings by using {\em EIS} spectral data
simultaneous to the selected {\em AIA} data, and establish the 
correspondence of the ``hot'' structures as singled out by the {\em AIA}
3 color image with the structures with bright \caxvii\ emission.
We devised a color coding that clearly emphasizes the hot structures.
Because of the full disk and high cadence nature of the {\em AIA} observations,
one of the most notable implications of our findings is the capability
of {\em AIA} to diagnose hot plasma at high spatial and temporal resolution,
therefore determining the degree of its spatial variability and temporal
evolution on timescales ranging from 12s, which is the typical cadence
of standard {\em AIA} observations, to roughly two weeks, which is the crossing
time from E to W limb.

\begin{acknowledgements}
We thank H.\ Warren for providing fitting routines for the {\em EIS} analysis,
not included in the SolarSoft package.
PT was supported by contract SP02H1701R from Lockheed-Martin, NASA
contract NNM07AB07C to the Smithsonian Astrophysical Observatory, and
NASA grants NNX10AF29G and 3001762433. 
FR acknowledges support from Italian Ministero dell'Universit\`a e 
Ricerca and Agenzia Spaziale Italiana (ASI) contract ASI/INAF I/023/09/0
``Attivit\`a scientifica per l'analisi dati Sole e plasma - Fase E2/F''.
\hinode\ is a Japanese mission developed and launched by ISAS/JAXA, with 
NAOJ as domestic partner and NASA and STFC (UK) as international 
partners. It is operated by these agencies in co-operation with ESA 
and the NSC (Norway).
\end{acknowledgements}

\end{document}